# SYNTHESIS AND SINTERING OF $MgB_2$ UNDER HIGH PRESSURE

T. A. Prikhna, Ya. M. Savchuk,
N. V. Sergienko, V.E. Moshchil,
S. N. Dub, and P. A. Nagorny
Institute for Superhard Materials
2 Avtozavodskaya St.
Kiev, 04074, Ukraine

W. Gawalek, T. Habisreuther,
M. Wendt, A. B. Surzhenko,
D. Litzkendorf, Ch. Schmidt and
J. Dellith
Institut für Physikalische
Hochtechnologie
Winzerlaer Str. 10
Jena, D-07743, Germany

V. S. Melnikov
Institute of Geochemistry,
Mineralogy and Ore-Formation
34 Palladin Pr.
Kiev, 03142, Ukraine

ABSTRACT

High-pressure (HP) synthesis and sintering are promising methods for manufacturing of the bulk $MgB_2$ superconductive material. The available high-pressure apparatuses with 100 cm$^3$ working volume can allow us to use the bulk $MgB_2$ for practical applications such as electromotors, fly-wheels, bearings, etc. We have found that the Ta presence during HP synthesis (especially) or sintering process (in the form of a foil that covered the sample and as an addition of Ta powder of about 2-10 wt.% to the starting mixture of B and Mg or to $MgB_2$ powder) increases the critical current density ($j_c$) in the magnetic fields up to 10 T and the fields of irreversibility ($H_{irr}$) of $MgB_2$-based bulk materials. We observed the strong evidences that Ta absorbs hydrogen and nitrogen during synthesis and sintering to form $Ta_2H$, $TaH$ and $TaN_{0.1}$ and prevents or reduces the formation of $MgH_2$ (both with orthorhombic and tetragonal structures). Ta also essentially reduces the amount of impurity nitrogen in black Mg-B (most likely, $MgB_2$) crystals of the matrix phase. The presence of Ta during synthesis allowed us to obtain a bulk materials with the following critical current densities $j_c$ in 1 T field: 570 kA/cm$^2$ at 10 K, 350 kA/cm$^2$ at 20 K and 40 kA/cm$^2$ at 30 K. Besides, the materials demonstrated high mechanical characteristics (microhardness, fracture toughness, Young modulus).

INTODUCTION

Magnesium diboride can be synthesized under ambient, elevated or high pressures. But the characteristics that are important for practical applications of a material as a superconductor, such as critical current density, irreversible magnetic field, etc. are very sensitive to the material density, impurity content and structural defects. As it was reported by A. Serquis et al.[1], the weak connectivity between domains and the presence of impurities in the grain boundaries in the MgB$_2$ could be the reason for the limited $j_c$ at high fields, although not as strong as in high-Tc superconductors. They are hot isostatically pressed (HIPed) at 200 MPa MgB$_2$ samples demonstrated the improved field dependence of $j_c$ through better connectivity of the grains, the generation of dislocations, and the destruction of MgO at MgB$_2$ boundaries, which were redistributed in the form of fine particles inside the MgB$_2$ matrix. These defects on authors' opinion can act as effective flux pinning centers. The HIPed samples had also a higher irreversibility field.

C.U. Jung et al.[2] after the investigation of a very pure single crystal of MgB$_2$ have suggested that the strong bulk pinning reported for polycrystalline material might be due to entirely extrinsic pinning sites, such as grain boundaries and crystallographic defects. This is consistent with the absence of core pinning, even at T~0.5xT$_c$, for bulk samples.

As a result of atomic resolution study of synthesized MgB$_2$ (experimental Z-contrast images and EEL spectra) R.F. Klie et al.[3] found ~20-100 nm sized precipitates that were formed by the ordered substitution of oxygen atoms onto boron lattice sites and that the basic bulk MgB$_2$ crystal structure and orientation were preserved. The periodicity of the oxygen ordering was dictated by the oxygen concentration in the precipitates and primarily occured in the (010) plane. The presence of these precipitates correlated well with an improved critical current density and superconducting transition behavior, implying that they act as pinning centers.

V.V. Flambaum et al.[4] investigating the hydrogenation of MgB$_2$ powder have found out that incorporation of hydrogen into MgB$_2$ structure can influence the superconductive properties (transition temperature) of the compound. They observed the increase in the transition temperature ($T_c$) of the $MgB_2H_x$ powder above that of the original powder, at least by ~ 0.5K for the $MgB_2H_{0.03}$ material. Authors suggest that these results confirm the assumption of a $T_c$ dependence on high frequency modes for an electron-phonon mediated superconductor. But whether hydrogenation can substantially increase the $T_c$ and the material, especially in the solid polycrystalline form, can be stable it's not studied yet.

The aim of the present investigations was to find the high-pressure-temperature conditions of synthesis (from Mg and B) and sintering (from $MgB_2$ powder) materials with high critical current density $j_c$ and irreversible field $H_{irr}$ and to study the correlations between the materials structure and properties. We continued also to study the positive effect of Ta presence during synthesis on superconductive properties of $MgB_2$ that was previously reported by us[5].

We succeeded in the synthesis of $MgB_2$-based materials with $j_c$ and $H_{irr}$ higher than those reported by Kijoon H. P. Kim et al.[6] and A. Serquis et al.[1] In Fig.1 we compared our data on the high pressure synthesized $MgB_2$ with the data on high-pressure sintered $MgB_2$ by Kijoon H. P. Kim et al.[6].

EXPERIMENTAL

In the experiments on synthesis, Mg and B have been taken in the stoichiometric ratio of $MgB_2$. To study the influence of Ta, the Ta powder has been added to the stoichiometric mixture of Mg and B powders in the amount of 2 and 10 wt. %. Then we mixed and milled the components in a high-speed activator with steel balls for 1-3 min. The obtained powder was compacted into tablets. In our experiments, we used two types of initial amorphous boron: type "A" – 95-97 % purity and type "B" – commercial boron produced about 25 years ago that during storing was partly transformed into $H_3BO_3$. For the experiments on sintering the commercial $MgB_2$ powder of Alfa Aesar company have been used.

The high-pressure conditions have been created inside a high-pressure apparatuses (HPA) of the recessed-anvil and cubic (six punches) types, described elsewhere.[7] Both types of apparatuses are usually used for diamond synthesis. They were slightly modernized for our purposes in order to measure temperature by thermocouples and to prevent the contact of the sample with the graphite heater. The working volume of the biggest cube-type HPA is of about 100 cm$^3$ (sample can be up to 60 mm in diameter). In our experiments, the sample was in contact with a compacted powder of hexagonal BN or enveloped in a Ta-foil and then placed inside the compacted BN or $ZrO_2$.

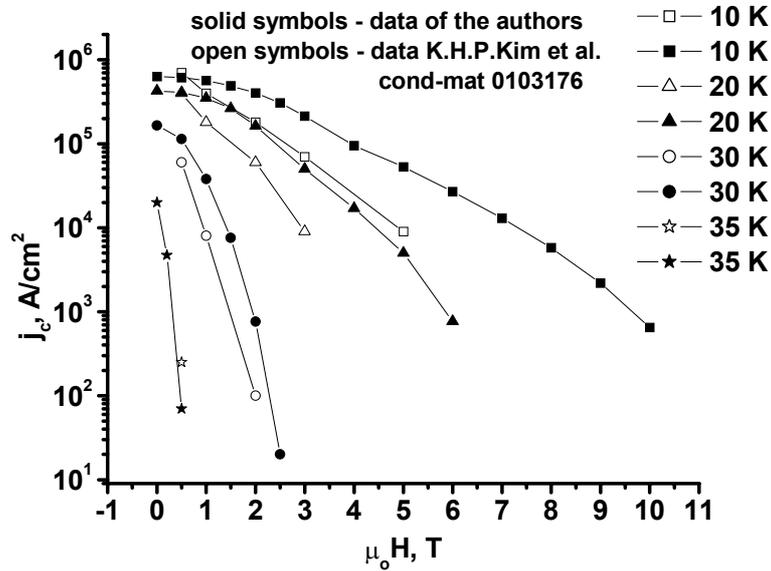

Fig.1. Data on critical current density $j_c$ vs. magnetic field $\mu H_o$ for the bulk $MgB_2$ samples:
solid symbols - high pressure synthesized at 2GPa, 800 – 900 $^\circ$C, 1h from Mg and B (of type "A") by the authors;
open symbols - high pressure sintered from $MgB_2$ powder at 3 GPa, 950$^\circ$C by Kijoon H. P. Kim et al.[6]

The structure of materials was studied using SEM and energy dispersive X-ray analysis, polarizing microscopy, X-ray structural and phase analysis. The $j_c$ was estimated from magnetization hysteresis loops obtained on an Oxford Instruments 3001 vibrating sample magnetometer (VSM) using Bean's model[8]. Hardness was measured on a Matsuzawa Mod. MXT-70 microhardness tester by a Vickers indenter. Nanohardness and Young modulus were investigated using Nano Indenter-II, MTS Systems Corporation, Oak Ridge, TN, USA. The fracture toughness was estimated from the length of the radial cracks emanating from the corners of an indent.

RESULTS AND DISCUSSION

Figure 2 shows VSM measurements of sintered $MgB_2$ powder. Figure 3 shows data of materials synthesized from magnesium and different types of boron. Figure 4 illustrates the structure of sintered (a) and synthesized (b) $MgB_2$ samples under a polarizing microscope; the structure of $MgB_2$ synthesized from boron type "A" in contact with BN (c) and with 10 wt.% of Ta addition (d) under a SEM.

Analysis of the obtained data allows us to conclude that the presence of Ta increases the critical current density ($j_c$) and irreversible field ($H_{irr}$) of the high pressure synthesized or sintered $MgB_2$-based material. Ta can be present during synthesis in the form of a foil that covers the sample or as addition to the starting mixture of B and Mg or to $MgB_2$ powder.

As X-ray, SEM and VSM studies show the superconductive properties ($j_c$, $H_{irr}$) of the materials are strongly influenced by the impurity content (of oxygen, hydrogen, nitrogen etc.) in the initial boron or magnesium diboride.

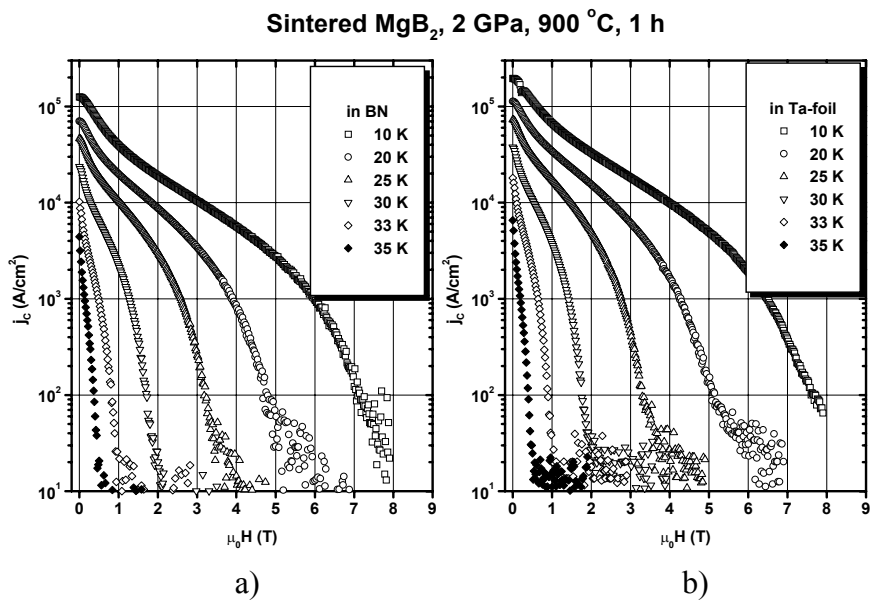

Fig.2 Critical current density $j_c$ vs. magnetic field $\mu_o H$ for the high pressure sintered $MgB_2$ samples at 2 GPa, 900 $^o$C, 1 h in BN (a) and in Ta-foil (b).

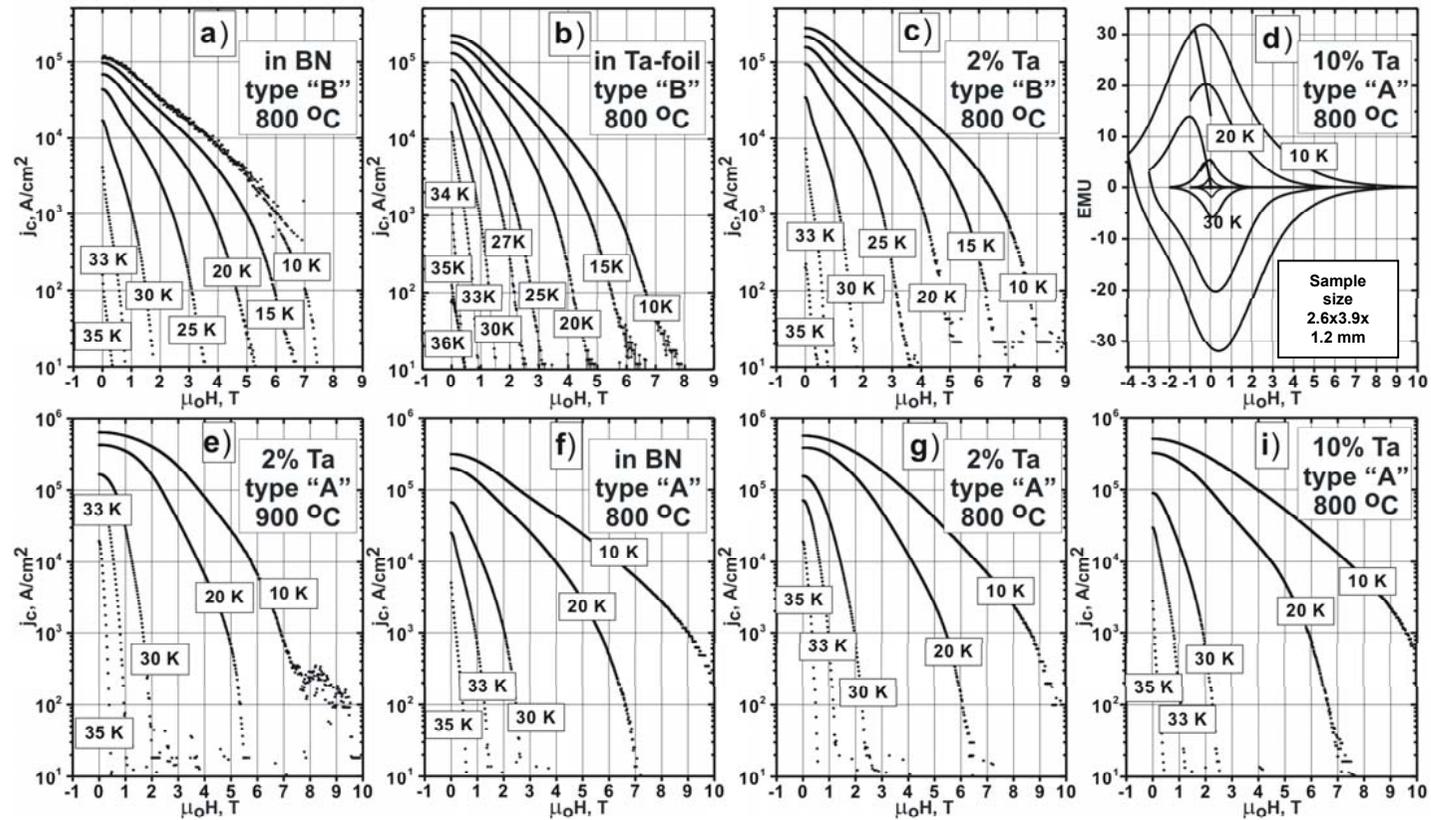

Fig. 3. VSM study of $j_c$ vs. $\mu_oH$ for the MgB$_2$ synthesized under 2 GPa, 1h:
(a), (b), (c) – using type "B" boron at 800 °C: (a) in contact with BN, (b) Ta-foil and (c) with 2 wt.% of Ta;
(e), (f), (g), (i)– using type "A" boron: (e) at 900 °C, with 2 wt.% of Ta, (f) at 800 °C in BN, (g) at 800 °C with 2 wt.% of Ta and (i) at 800 °C with 10wt.% of Ta; Fig.3d - magnetization loops of the sample shown on Fig.3i.

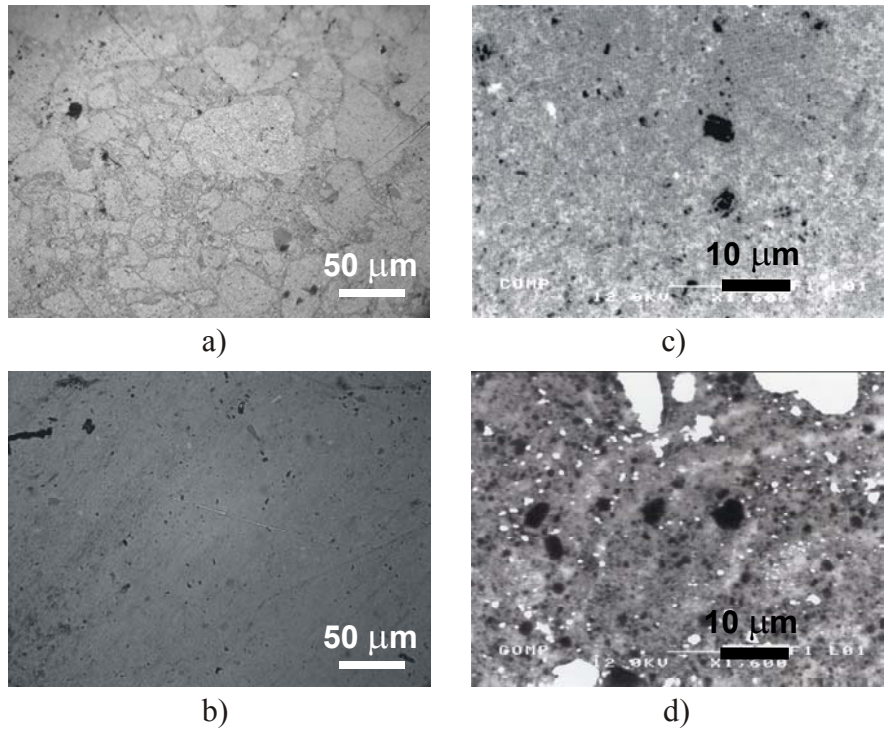

Fig.4. Photos of the structures of MgB$_2$ obtained at 2 GPa, 900 °C, 1 h by sintering of MgB$_2$ powder (a) and by synthesis from Mg and B (type "A") with 2 wt.% of Ta (b) taken under a polarizing microscope;

SEM pictures (composition image) that show the different concentrations of black Mg-B (evidently, MgB$_2$) grains in the samples synthesized from magnesium and type "A" boron at 2 GPa, 800 °C, 1h: (c) in contact with BN (without Ta presence) and (d) with 10 wt.% of Ta addition.

High-pressure synthesized and sintered samples have a multiphase nanostructure. As SEM study shows the matrix phase of the samples consists mainly of Mg, B, O. The black grains or single crystal inclusions (from the micron or even less to dozen microns in size) of Mg-B phase (MgB$_2$) are distributed in the matrix. The distribution of black MgB$_2$ grains in the matrix in high-pressure synthesized material is more homogeneous than that in the sintered one (Fig.4 a, b).

We observed the strong evidence that Ta absorbs gases during the manufacturing process:

a) $Ta_2H$ and $TaH$ were detected after the synthesis and $TaN_{0.1}$ was detected after the sintering in the materials, while no Ta-Mg, Ta-B or Ta-Mg-B compounds have been found;

b) we haven't observed the $MgH_2$ presence either with orthorhombic or tetragonal structure in the samples synthesized from boron "A" (without impurity of $H_3BO_3$) into which Ta (2 - 10 wt.%) was added. While in the samples synthesized without Ta addition the $MgH_2$ (both structures) were found. When boron "B" (commercial, with a high amount of impurity $H_3BO_3$) have been used as starting material, the addition of about 20 wt.% of Ta only allowed us to avoid the $MgH_2$ formation;

c) black inclusions (or single crystals) of $MgB_2$ phase in the samples synthesized from "A" amorphous boron with Ta addition contain no impurity nitrogen and less impurity oxygen than those in the samples synthesized without Ta addition.

The samples with higher $j_c$ and $H_{irr}$ have the higher density of Mg-B ($MgB_2$) inclusions in their matrix, i.e. the higher amount of black grains (see, for example, Figs. 4 c, d). Besides, in the samples with better superconductive properties, these black Mg-B inclusions contain a higher amount of boron than that in the samples with worse superconductive characteristics while the amount of magnesium is about the same. The samples with better superconductive properties also contained a higher amount of unreacted magnesium in the matrix phase.

Positive influences of Ta are much more pronounced in the synthesis process than in the sintering one. The presence of Ta extends the temperature region of synthesis of a material with high $j_c$ and $H_{irr}$.

The results of investigations of mechanical properties are given in Table 1. The black inclusions ($MgB_2$ single crystals) have hardness higher than that of sapphire.

CONCLUSIONS

The positive effect of Ta presence during synthesis of $MgB_2$ on critical current density and field of irreversibility (evidences of which were shown by us earlier) has been confirmed. Ta does not react with Mg or B during the manufacturing process but absorbs gases: hydrogen, nitrogen, etc. By adding Ta and using amorphous boron free from $H_3BO_3$ impurity, the $MgB_2$-based material with high critical current density at temperatures 10 - 30 K in magnetic fields up to 2-10 T have been obtained. We observed a lot of differences in the structure of materials with different superconductive properties and it seems that for the higher values of $j_c$ and $H_{irr}$, the higher amount of black $MgB_2$ grains in the matrix phase (1) as well as the lower content of nitrogen and oxygen but higher boron in them (2), more uniform distribution of these black grains in the matrix (3), presence of some amount of unreacted magnesium (4) can be responsible. May be the important role played milling and mixing of the initial powders by the high-speed activator (5) and our worst results in high –pressure sintering are due to rather big sizes of grains of the starting $MgB_2$ powder. Because, as the polarizing microscope shows, the density of black grains in the sintered material is higher in the places of former boundaries between the initial particles, thus black grains seems to "repeat" the structure of the initial powder (Figs. 4 a, b).

Table 1. Results of micro- (Vickers indent) and nanohardness (Berkovich indent) investigations of the sintered $MgB_2$ samples

| Characteristics | Matrix phase of the samples | Single crystal $MgB_2$ | Sapphire, $Al_2O_3$ |
|---|---|---|---|
| Indentation load: 60 mN | | | |
| Nanohardnes, $H_B$, GPa | 17.4±1.1 | 35.6±0.9 | 31.1±2.0 |
| Young modulus, E, GPa | 213±18 | 385±14 | 416±22 |
| Indentation load: 4.96 N | | | |
| Vickers microhardness, $H_v$, GPa | 17.1±1.11 | - | - |
| Indentation load: 147.2 N | | | |
| Fracture toughness, $K_{1c}$, MN/m$^{-3/2}$ | 7.6±2.0 | - | - |
| Vickers microhardness, $H_v$, GPa | 10.12±0.2 | - | - |

The hardness of the black single crystalline $MgB_2$ grains turned out to be higher than that of sapphire single crystals.

The attained level of superconductive and mechanical properties of the high-pressure synthesized and sintered $MgB_2$ and the possibility to produce large bulk $MgB_2$ products make this material very promising for practical applications.

ACKNOWLEDGEMENTS
We are grateful to Prof. S.Abell (University of Birmingham) for the support of the present investigations.

REFERENCES

[1] A. Serquis, X.Y. Liao, Y.T. Zhu, J.Y. Coulter, J.Y. Huang, J O. Willis, D.E. Peterson, F.M. Mueller, N.O. Moreno J.D. Thompson, S.S. Indrakanti and V.F. Nesterenko, "The influence of microstructures and crystalline defects on the superconductivity of $MgB_2$", *cond-mat/*0201486.

[2] C.U. Jung, J.Y. Kim, P. Chowdhury, Kijoon H.P. Kim and Sung-Ik Lee, D.S. Koh, N. Tamura, W.A. Caldwell and J.R. Patel, "Microstructure and pinning properties of hexagonal-disc shap ed single crystalline $MgB_2$", *cond-mat/*0203123.

[3] R.F. Klie, J.C. Idrobo, N. D. Browning, A.C. Serquis, Y.T. Zhu, X.Z. Liao and F.M. Mueller, "Observation of coherent oxide precipitates in polycrystalline $MgB_2$", *cond-mat/*0203292.

[4] V.V. Flambaum, G.A. Stewart, G.J. Russell, J. Horvat and S.X. Dou, "Superconducting transition temperature of $MgB_2H_{0.03}$ is higher than that of MgB2", *cond-mat/*0112301.

[5] T.A. Prikhna, W. Gawalek, A. B. Surzhenko, N.V. Sergienko, V.E. Moshchil, T. Habisreuther, V. S. Melnikov, S.N. Dub, P.A. Nagorny, M. Wendt, Ya.M. Savchuk, D. Litzkendorf, J. Dellith, S. Kracunovska, Ch. Schmidt, "The high-pressure synthesis of $MgB_2$", *cond-mat/*0109216

[6] Kijoon H. P. Kim, W. N. Kang, Mun-Seog Kim, C.U. Jung, Hyeong-Jin Kim, Eun-Mi Choi, Min-Seok, Park & Sung-Ik Lee, "Origin of the high DC transport critical current density for the $MgB_2$ superconductor", *cond-mat/*0103176.

[7] T. Prikhna, W. Gawalek, V. Moshchil, S. Dub, T. Habisreuther, V. Melnikov, F. Sandiumenge, V. Kovylayev, A. Surzhenko, P. Nagorny, P. Schaetzle, A. Borimsky, "Improvement of properties of Y- and Nd-base melt textured high temperature superconductors by high pressure-high temperature treatment", p.p.153-158 in *Functional Materials,* Edited by K.Grassie, E. Teuckhoff, G.Wegner, J.Hausselt, H.Hanselka, EUROMAT99-V.13, 2000.

[8] C. B. Bean, "Magnetization of high-field superconductors," *Rev. Mod. Phys.*, **36** 31-36 (1964).